\documentclass[prd,superscriptaddress,twocolumn,10pt]{revtex4}

\usepackage{amsmath,amssymb}
\usepackage{verbatim}
\usepackage{graphicx}
\usepackage{hyperref}
\usepackage{color} 

\DeclareFontFamily{OT1}{rsfs}{}
\DeclareFontShape{OT1}{rsfs}{m}{n}{ <-7> rsfs5 <7-10> rsfs7 <10->rsfs10}{} 
\DeclareMathAlphabet{\mycal}{OT1}{rsfs}{m}{n} 

\begin{document}

\title{Coupling coefficient in three dimensional higher spin holography }

\author{Iva Lovrekovic}
\email{ilovreko@.ic.ac.uk}
\affiliation{The Blackett Laboratory, Imperial College London,
Prince Consort Road, London SW7 2AZ, United Kingdom}

\date{\today}


\begin{abstract}We consider linearized Vasiliev equations around the background AdS
field in three dimensions for the correlation function of two scalars and a higher spin field. Relating this with the higher spin field determined in the metric formulation allows determination of the corresponding coupling coefficient. The result agrees with the analogous computation for the spin three field.
\end{abstract}
\maketitle

\section{Introduction and Motivation}

Higher spin (HS) theories have started receiving more attention in the recent years due to a fact that they offer an answer to some important questions in string theory. They have been introduced in the early years by Vasiliev  \cite{Fradkin:1986qy,Vasiliev:1990en,Fradkin:1987ks,Vasiliev:1995dn}. We can describe them characterizing free fields by spin and mass and considering the consistent interaction among them. Writing consistent interaction for fields of arbitrary spin and mass turned out to be difficult, therefore one usually starts by looking at the massless fields. Their particular property is that interaction terms due to the masslessness need to be restricted by gauge symmetry, making them interesting and simpler to study. For the fields with spin higher than two and the general number of dimensions higher than three, the spectrum of the theory is necessarily infinite and contains the fields of all spins. In three dimensions it is possible to truncate to a finite number and one obtains the Chern-Simons action which describes a coupling of massles HS fields and AdS gravity in consistent manner. The gauge group is two copies of $SL(n,\mathbb{R})$ \cite{Blencowe:1988gj,Bergshoeff:1989ns,Henneaux:2010xg} and for finite n one can consider finite number of interacting HS fields \cite{Campoleoni:2010zq}. The theory is minimal, which means that each spin state appears only once, unlike in the string theory, where spectrum contains degenerate states. Another property of the theory is that it does not contain dimensional parameter like string theory therefore it cannot come from spontaneous symmetry breaking.

In the early days of its development the theory encountered number of no-go theorems due to a fact that interacting higher-spins do not propagate in the Minkowski space. This issue was solved by consideration of the HS fields on AdS.

First action for arbitrary HS fields was introduced by Fronsdal and described a tower of massless noninteracting HS fields \cite{Fronsdal:1978rb,Fronsdal:1978vb,Fang:1978wz}. The issue that remained is construction of the interacting action of the theory for the spin higher than three. The interacting theories can be constructed order by order using perturbative interaction procedure. One deforms a quadratic Lagrangean of a particle spectrum of certain spins and masses by cubic terms, while keeping the gauge invariance which he repeats at the following step for quartic terms, etc. The procedure also deforms the gauge transformations. The solution of the cubic deformations gives cubic interaction vertices. The cubic vertices have been studied from the metriclike and framelike approach in number of articles. Some of the early works in metriclike formalism are \cite{Berends:1984rq,Metsaev:1993mj}\footnote{For the list of references see a review on Vasiliev equations \cite{Didenko:2014dwa}.}.These traditional methods as metriclike approach have led to little progress in development of construction of interacting theory for higher-spin gauge fields. They are less efficient beyond cubic interaction.

The new approach to this was offered by Fradkin and Vasiliev, and extended by Vasiliev, developing nonlinear system for higher spin fields, called unfolded approach \cite{Vasiliev:1990en,Vasiliev:1990cm}. An overview of this approach can be found in \cite{Didenko:2014dwa}, which also considers Vasiliev equations up to second order in perturbation. First attempt to extract observables from the equations has been done in \cite{Sezgin:2003pt}, while the extraction of the higher spin couplings from Vasiliev equations has been considered in \cite{Giombi:2009wh}.
 
HS theory can also be considered in the sense of AdS/CFT correspondence, where the HS theory of massless HS fields corresponds to a limiting case of the string theory for the string tension going to zero. Large N superconformal field theories were studied as holographic duals for higher spin gauge theories in perturbative expansion around AdS spacetime \cite{Sundborg:2000wp}.

Klebanov and Polyakov \cite{Klebanov:2002ja} have proposed a duality  between the singlet sector of the critical 3-d O(N) vector model with $(\phi^a\phi^a)^2$ interaction and minimal bosonic theory in AdS4 which contains massless gauge fields with even spin. The analog of AdS4 conjecture by Klebanov and Polyakov appeared in AdS3 \cite{Gaberdiel:2010pz} conjecturing a duality between a complex scalar coupled to higher-spin fields in Vasiliev's gravity in 3 dimensions and WN minimal model CFT in t'Hooft limit denoted by coset representation
 \begin{equation}
\frac{SU(N)_k\oplus SU(N)_1}{SU(N)_{k+1}},
\end{equation}
where we define the t'Hooft limit with $N,k\rightarrow\infty$ for $\lambda\equiv\frac{N}{k+N}$. 
This duality has been verified by the number of studies, correspondence of global symmetries in bulk and at the boundary \cite{Gaberdiel:2011wb}, correspondence of the 1-loop partition function in the bulk and at the large NCFT  \cite{Gaberdiel:2011zw}, and partition function of the HS black hole at high temperature in the bulk and at the boundary CFT, as well as for the 3-point functions when $\lambda=\frac{1}{2}$ and $s=2,3,4$ for the scalar-scalar-HS field (00s) correlator in the t'Hooft limit. 
Via three-point functions, tests of the conjecture have been done in \cite{Ammon:2011ua} (for 00s correlator with general $\lambda$), and in \cite{Chang:2011mz,Skvortsov:2015lja}.

In this work, we extract the coupling of the 00s  three-point correlator  by considering the linearised Vasiliev equations of motion, and we verify it by choosing the spin to be three and comparing with result in \cite{Ammon:2011ua}.
The result corresponds to coupling of the three-point correlation function up to selected normalisation. 
While we consider general $\lambda$, the similar work has been done for the fixed $\lambda$ in \cite{Kessel:2015kna}.

The structure of the work is as follows: In the section two we consider linearised equations of motion in the Vasiliev's theory, in the section three we consider the higher spin field in the metric formulation, and in section four we conclude.

  \section{Linearized equations of motion}

Let us first consider the coefficient coming from the Vasiliev linearised equations. Vasiliev's theory contains five equations for the master fields W which is spacetime 1-form, B and $S_{\alpha}$ which are spacetime 0-forms. The generating functions are dependent on the coordinates of the spacetime, auxiliary bosonic twistor variables (referred to as "oscillators") and Clifford element pairs, where in definitions we follow conventions from \cite{Ammon:2011ua}. 
The oscillators  and various other ingredients are used to define the "deformed" oscillator star-commutation relations which give rise to $hs[\lambda]$ higher spin algebra. 
Two of the above mentioned equations that will be of the interest here are
\begin{align}
dW&=W\wedge\star W \\
dB&=W\star B-B\star W 
\end{align}
We can rewrite W with projector operators 
\begin{equation}
\mathcal{P}_{\pm}=\frac{1\pm\psi}{2} 
\end{equation}
for $\psi$ elements of the Clifford pairs such that $W=-\mathcal{P}_{+}A-\mathcal{P}_{-}\overline{A}$
for 
\begin{align}
\mathcal{P}_{\pm}\psi_{1}&=\psi_1\mathcal{P}_{\pm}=\pm\mathcal{P}_{\pm} &\mathcal{P}_{\pm}\psi_2=\psi_2\mathcal{P}_{\mp}
\end{align}
where A are Chern-Simons gauge fields which take value in the Lie algebra hs$[\lambda]$. In this formulation the equation 
\begin{equation}
dW=W\wedge\star W
 \end{equation}
gives
\begin{align}
dA+A\wedge\star A =0 \label{da}\\ 
D\overline{A}+\overline{A}\wedge \star \overline{A}=0 \label{dabar}
\end{align}
where A and $\overline{A}$ are positive polynomials of the positive degree in products of deformed oscillators. 
(\ref{da}) and (\ref{dabar}) are in that case equal to field equations $hs[\lambda]\otimes hs[\lambda]$  Chern-Simons theory. \\ 
The generators of hs$[\lambda]$ are defined with spin index \textit{s} and mode index \textit{m}  as
$V^s_{m}$ for $s\geq2 $
while $|m|<s$ and obey 
 star product
\begin{equation}
V_m^s\star V_n^t=\sum_{u=1,2,3}^{s+t-|s-t|-1}g_u^{st}(m,n;\lambda)V^{s+t-u}_{m-n}\label{starproduct}
\end{equation}
where 
\begin{equation}g_u^{st}(m,n;\lambda)=(-1)^{u+1}g_u^{ts}(m,n;\lambda)\end{equation}
are specific coefficients dependent on $\lambda$ and defined according to conventions \cite{Ammon:2011ua}.
The equations describe interaction of arbitrary higher spin background with lienarized scalars. 
The coupling that we are interested in can be extracted from rewriting the master field B as a linearized fluctuation around vacuum value $\mathcal{\nu}$
\begin{align}
B=\mathcal{\nu}+\mathcal{P}_{+}\psi_2C(x,\tilde{y}_{\alpha})+\mathcal{P}_-\psi_2\tilde{C}(x,\tilde{y}_{\alpha})
\end{align}
and expanding the master field $C$ in the deformed oscillators $\tilde{y}_{\alpha}$ in the equation
\begin{equation}
dC+A\star C- C \star \overline{A}=0.\label{ceqn}
\end{equation}
 That allows us determining the generalised Klein-Gordon (KG) equation in the background of HS fields. 
While the expansion of the master field C in formalism of bosonic Vasiliev's theory is given by
 \begin{equation}
 C=C_0^1+C^{\alpha\beta}\tilde{y}_{\alpha}\tilde{y}_{\beta}+C^{\alpha\beta\sigma\lambda}\tilde{y}_{\alpha}\tilde{y}_{\beta}\tilde{y}_{\sigma}\tilde{y}_{\lambda}+...
 \end{equation}
 with implied star product and symmetric C components.
  Master field components are now separated in physical scalar field $C_0^1$  and higher ones, related to it on-shell by derivatives. 
The expansion of C is given by
\begin{equation}
C=\sum_{s=1}^{\infty}\sum_{|m|<s}C_m^sV_m^s
\end{equation}
for $C_m^s\sim C^{\alpha_1\alpha_2..\alpha_{2s-1}}$
where m and number of oscillators $\tilde{y}_{1}$ versus $\tilde{y}_2$ are related with $2m=N_1-N_2$ and $C_m^s$ are functions of spacetime coordinates. Auxiliary tensors are absorbed within a field.
The fields $A$ and $\bar{A}$ are expanded analogously
\begin{align}
A&=\sum_{s=2}^{\infty}\sum_{|m|<s}A_m^sV_m^s & \overline{A}=\sum_{s=2}^{\infty}\sum_{|m|<s}\overline{A}_m^sV_m^s.
\end{align}
The standard procedure of finding the generalised KG equation consists of inserting the expressions for $A$, $\bar{A}$ and $C$ in  (\ref{ceqn}) and determining the smallest possible set of equations needed to find the scalar equation in arbitrary background.
Standard procedure can be described considering equation (\ref{ceqn}) in AdS background since it is a foundation for the following computations. The vacuum $C_0^1$ equation without AdS fields is ordinary KG  equation while one can determine the higher components in the terms of $C_0^1$. 

The AdS connection consists of the spin-2 generators that form SL(2), subalgebra of $hs[\lambda]$ 
\begin{align}
A&=e^{\rho}V_1^2dz+V_0^2d\rho\\
\bar{A}&=e^{\rho}V_{-1}^2d\bar{z}-V_0^2d\rho
\end{align}
with AdS metric
\begin{equation}
ds^2=d\rho^2+e^{2\rho}dzd\bar{z}. \label{adsmetric}
\end{equation}
The higher spins fields vanish, and we are working in Euclidean metric and  Fefferman-Graham gauge. The general form of the $C$ equation (\ref{ceqn}) in the AdS background is 
\begin{align}
&\partial_{\rho}C_m^s+2C_m^{s+1}+C_m^{s+1}g_3^{(s+1)2}(m,0)=0  \label{gencads1}\end{align}
\begin{align}
 \partial C_m^s+e^{\rho}(C_{m-1}^{s-1}+\frac{1}{2}g_2^{2s}(1,m-1)C^s_m\\ \nonumber+\frac{1}{2}g_3^{2(s+1)}(1,m-1)C_{m-1}^{s+1})=0 \label{gencads2}\end{align}
\begin{align}
\overline{\partial}C^s_m-e^{\rho}( C_{m+1}^{s-1}-\frac{1}{2}g_2^{2s}(-1,m+1)C^s_{m+1}\\+\frac{1}{2}g_3^{2(s+1)}(-1,m+1)C^{s+1}_{m+1})=0 \nonumber\label{gencads3}
\end{align}
for $|m|<s$, $\partial=\partial_z,\overline{\partial}=\partial_{\overline{z}}$ and the $\lambda$-dependence in the structure constants suppressed. 

In the simplest case choosing $s=1,s=2$ one can solve for the higher components in C and obtain the Klein-Gordon KG equation 
\begin{equation}
\left[ \partial_{\rho}^2+2\partial_{\rho}+4e^{-2\rho}\partial\bar{\partial}-(\lambda^2-1) \right]C_0^1=0\label{kgeqn}.
\end{equation}
 Consistency condition on equations is that all the components of C have smooth solution when expressed using $C_0^1$. The strategy for determining the minimal set of equations for $C_0^1$ is to select components of C that are of the form $C_{\pm m}^{m+1}$ and therefore the smallest spin for fixed m (e.g. $C_0^1, C_{\pm1}^2,..$). That are minimal components. 
 One needs $V_{m,\rho}^s$ equations for fixed m, solve for non-minimal components in terms of minimal ones and $\rho$ derivatives, for $A_{\rho}=-\overline{A}_{\rho}=V_{0}^2$. After solving for minimal ones, one needs to solve $V_{m,z}^s$ and $V_{m,\overline{z}}^s$ equations in terms of $C_0^1$ and its derivatives. 
 
Once that we have expressed the higher components of C in terms of $C_0^1$ we can determine the part that defines the KG equation and the
generalised part that appears due to the HS background.
To obtain the equation of motion for the scalar field up to linear order we consider the variation of the gauge field and apply the KG equation on it. This and the standard procedure for obtaining the linearised equation of motion for the scalar field described above 
should be equal once the gauge parameter is chosen conveniently.
That approach can be written in the following way.

First we express the higher components of the C field in terms of the combination of the derivatives on $C_0^1$ in the background AdS. 
Focusing on the master field $C$, the equation (\ref{ceqn}) is invariant under the $hs[\lambda]\oplus hs[\lambda]$  gauge invariance when 
\begin{align}
C\rightarrow C+C\star \bar{\Lambda}-\Lambda\star C \label{invc}
\end{align}
for 
\begin{equation}
\Lambda(\rho,z,\bar{z})=\sum_{n=1}^{2s-1}\frac{1}{(n-1)!}(-\partial)^{n-1}\lambda^{(s)}(z,\bar{z})e^{(s-n)\rho}V^{s}_{s-n}. \label{gaugeinv}
\end{equation}
Where we take $\Lambda$ to be chiral, so $\bar{\Lambda}=0$.
The field in the higher spin background is obtained by transformation
\begin{align}
\tilde{C}_m^s=C_m^s-(\Lambda\star C)^s_m.
\end{align}
The field $C_m^s$ we express in terms of the $C_0^1$.
To do that we focus on the set of equations (\ref{gencads1}),(20),(\ref{gencads2}).
The product of the C field with $\Lambda$ gives combination of higher components of C in AdS background which can, as we will show, be expressed in terms of $C_0^1$. On the field $C_0^1$ we can use the transformation (\ref{invc}) and obtain 
\begin{align}
\tilde{C}_0^1=C_0^1-(\Lambda\star C)^1_0.
\end{align}
Since we are at the linear order, once we have $C_0^1$ we can rewrite it as $\tilde{C}_0^1$ which is defined on the higher spin background. 
From the expression for the gauge field $\Lambda$ (\ref{gaugeinv}) and the relation for the star product (\ref{starproduct}) we can determine the variation of the scalar field $C^1_0$
\begin{align}
(\delta C)^1_0 &=-\sum_{n=1}^{2s-1}\frac{1}{(n-1)!}(-\partial)^{n-1}\Lambda^{(s)} \\ &\times\frac{1}{2}g_{2s-1}^{ss}(s-n,n-s)C^s_{-(s-n)}e^{(s-n)\rho}, \label{varc1}
\end{align}
for $C^s_{-(s-n)}$ an arbitrary component of the master field $C$.
Taking $m\rightarrow -m$ in the set of equations (\ref{gencads1}), (20), (\ref{gencads2}) and $s=m+1$ in (\ref{gencads2}), we can iteratively determine the dependence of the $C^{m+1}_m$ on the $C_0^1$. From the equation (\ref{gencads2}) we obtain
\begin{equation}
\partial_z C_{-m}^{m+1}+\frac{e^{\rho}}{2}g_3^{2(m+2)}(1,-m-1)C^{m+2}_{-m-1}=0\label{citerat}
\end{equation}
taking into consideration that for certain components $C_n^s$ it is required $|n|\leq s-1$ this iteratively leads to relation of $C_m^{m+1}$ and $C_0^1$, and from  
the (\ref{gencads2}) analogously for $C^{m+1}_{-m}$ and $C_0^1$.
The general form of the $C^s_{\pm}$ is 
then given in terms of $C^{m+1}_{\pm m}$ and coefficients $g_u^{ts}(m,n)$.
Knowing $C^s_{\pm}$ and $C^{m+1}_{\pm m}$ allows to obtain  \cite{Ammon:2011ua}
\begin{align} 
(\delta C)^1_0&=\sum_{n=1}^sf_{\pm}^{s,n}(\lambda)\partial_z^{n-1}\Lambda^{(s)}\partial_z^{s-n}\phi \label{varc2}
\end{align}
for $\phi\equiv C_0^1$ and $f^{s,n}_{\pm}(\lambda)$ expressed in terms  of coefficients $g_u^{st}(m,n)$
Using the replacement $\partial_{\rho}\rightarrow-(1\pm\lambda)$ 
and 
 writing explicitly first few n values for $f_{\pm}^{s,n}(\lambda)$, allows to determine its general expression 
\begin{align}
f_{\pm}^{s,n}(\lambda)&=(-1)^s\frac{\Gamma(s+\lambda)}{\Gamma(s-n+1\pm\lambda)}\frac{1}{2^{n-1}(2(\frac{n}{2}-1))!!\left(\frac{n-1}{2}\right)!}\nonumber\\
&\times\prod_{j=1}^{\frac{n-1}{2}}\frac{s+1-n}{2s-2j-1} \label{fpm}.
\end{align}
\noindent Substituting (\ref{fpm}) in (\ref{varc2}) one obtains the variation of the scalar field 
\small
\begin{align}
\displaystyle{(\delta C)_0^1}&\displaystyle{=\sum_{n=1}^{s}(-1)^s\frac{\Gamma(s\pm\lambda)}{\Gamma(s-n+1\pm\lambda)}\frac{1}{2^{n-1}\left(2\left(\frac{n}{2}\right)-1\right)!!\left(\frac{n-1}{2}\right)!} \nonumber}\\ & \times\prod_{j=1}^{\left( \frac{n-1}{2} \right)}\frac{s+j-n}{2s-2j-1}\partial_z^{n-1}\Lambda^{(s)}\partial_z^{s-n}C_0^1. \label{eomvar}
\end{align}
\normalsize
To consider the coefficient in front, we focus on the term with the lowest number of $\partial_z$ derivatives on the gauge field $\Lambda^{(s)}$,  obtained for n=1. Then, (\ref{eomvar}) becomes
\begin{equation}
(\delta C)_0^1|_{n=1}=(-1)^s\Lambda^{(s)}\partial^{s-1}C_0^1 \label{varc01}.
\end{equation}
To obtain the linearised equation of motion for the scalar field we act on (\ref{varc01}) with KG operator (\ref{kgeqn}).
This can be written as
\begin{equation}
\Box_{KG}\tilde{C}^1_0=\Box_{KG}C_0^1+\Box_{KG}\delta C_0^1 \label{boxonvar}.
\end{equation}
Taking   $\partial_{\rho}\rightarrow(1\pm\lambda)$ in $f_{\pm}^{s,n}(\lambda)$ we have taken and considering the term with highest number of derivatives on $C_0^1$  leads to
\begin{align}
&\Box_{KG}|_{\text{highest number of derivatives}}(\delta C)_0^1=\\& = (-1)^s4e^{-2\rho}\partial(\bar{\partial}\Lambda^{(s)}\partial^{(s-1)}C_0^1)\\&=(-1)^s4e^{-2\rho}[ \partial\bar{\partial}\bar{\Lambda}^{(s)}\partial^{(s-1)}C_0^1+\bar{\partial}\Lambda^{(s)}\partial^{s}C_0^1\nonumber\\&+\partial\Lambda^{(s)}\bar{\partial}\partial^{(s-1)}C_0^1+\Lambda^{(s)}\bar{\partial}\partial^sC_0^1 ]\label{boxeom}.
\end{align}
The term in (\ref{boxeom}) that is of further interest is the one multiplying $4e^{-2\rho}\partial\bar{\partial}$ acting on $\delta C_0^1$ which is convenient to compute in the metric formulation.

\section{Metric formulation}
In the metric formulation we can express the higher spin field of arbitrary spin $s$ with
\begin{equation}
\phi_{\mu_1.....\mu_s}=tr\left( \tilde{e}_{(\mu_1}...\tilde{e}_{\mu_{s-1}}\tilde{E}_{\mu_s)} \right)\label{metricform}
\end{equation}
where $\tilde{E}_{\mu s}=\tilde{A}_{\mu}-\tilde{\bar{A}}_{\mu}$ and $\tilde{A}_{\mu}$ and $\tilde{\bar{A}}_{\mu}$ we define below. 
The  dreibein is determined from the background AdS metric (\ref{adsmetric}) 
\begin{align}
e_{z}&=\frac{1}{2}e^{\rho}(L_1+L_{-1})=\frac{1}{2}e^{\rho}(V_1^2+V_{-1}^{2}) \\
e_{\bar{z}}&=\frac{1}{2}e^{\rho}(L_1-L_{-1})=\frac{1}{2}e^{\rho}(V_1^2-V_{-1}^2) \\
e_{\rho}&=L_0=V_0^2.
\end{align}
The invariance of the equation (\ref{ceqn}) under the gauge transformation for $hs[\lambda]\oplus hs[\lambda]$ for the fields A means 
\begin{align}
A&\rightarrow A+d \Lambda +\left[A,\Lambda \right]_{\star}\equiv \tilde{A}\\
\bar{A}&\rightarrow \bar{A}+d \bar{\Lambda} +\left[\bar{A},\bar{\Lambda} \right]_{\star}\equiv \tilde{\bar{A}}.
\end{align}
Since $\Lambda$ parameter is chiral it means $\bar{\Lambda}=0$ and the field $\tilde{\bar{A}}$ is essentially unchanged. The field $\tilde{A}_{\mu}$ is then  
\begin{equation}
\tilde{A}=A_{AdS}+d\Lambda+\left[A_{AdS},\Lambda\right]_{\star}.
\end{equation}
 $d\Lambda$  reads
\small
\begin{align}
d\Lambda&=\sum_{n=1}^{2s-1}\frac{1}{(n-1)!}V_{s-n}^se^{(s-n)\rho}[ (-\partial)^{n-1}\partial\Lambda^{(s)}(z,\bar{z}) dz\\&+(-\partial)^{n-1}\bar{\partial}\Lambda^{(s)}(z,\bar{z})d\bar{z}+(-\partial)^{n-1}\Lambda^{(s)}(z,\bar{z})(s-n)d\rho ] \label{dlambda}
\end{align}
\normalsize
and 
\begin{align}
\left[A_{AdS},\Lambda\right]_{\star}&= [ e^{\rho}V_1^2dz+V_0^2d\rho,\nonumber\\ &\sum_{n=1}^{2s-1}\frac{1}{(n-1)!}(-\partial)^{n-1}\Lambda^{(s)}(z,\bar{z})e^{(s-n)\rho}V^s_{s-n} ]
\end{align}
To read out the coupling we focus on $\bar{z}....\bar{z}$ component of the field $C_0^1$ with lowest number of derivatives on gauge field $\Lambda^{(s)}$. The $\star$ multiplication of the dreibeins in (\ref{metricform}) in that case contributes only with first $g_{u}^{st}(m,n;\lambda)$  coefficient with the each following dreibein that is being multiplied. More explicitly 
\begin{align}
e_{\bar{z}}\star e_{\bar{z}}&=\frac{1}{2^2}e^{2\rho}\left(V_1^2-V_{-1}^2\right)\star(V_{1}^2-V_{-1}^2)
\end{align}
From (\ref{gaugeinv}) we notice that the lowest number of derivatives on $\Lambda$ will appear for lowest n, i.e.  for $n=1$ in summation (\ref{gaugeinv}). Knowing the relation for the trace of higher spin generators, the required generator $V^s_{s-n}$ will than be of the form $V^s_{s-1}$, as we see below, which means that multiplication of HS generators we have to  consider is 
\begin{equation}
V_{-1}^2\star V_{-1}^2\star....\star V_{-1}^2.
\end{equation}
Then 
\begin{align}
V_{-1}^2\star V_{-1}^2 &=\frac{1}{2}( g_1^{22}(-1,-1)V_{-2}^3+g_{2}^{22}(-1,-1)V_{-2}^2\nonumber\\&+g_{3}^{22}(-1,-1)V_{-2}^1 )
\end{align}
where the $g_2^{22}(-1,-1)=g_3^{22}(-1,-1)=0$. Multiplying with following $V_{-1}^2$, etc. on e can conclude
\begin{equation}
\underbrace{V_{-1}^{2}\star V_{-1}^2\star ....\star V_{-1}^2}_{s-1}=\frac{1}{2^{s-1}}g_1^{2(s-1)}(-1,-(s-2))V_{-(s-1)}^s
\end{equation}
while 
\begin{equation}
g_{2}^{2(s-1)}(-1,-(s-2))=g_{3}^{2(s-1)}(-1,-(s-2))=0.
\end{equation}
That  means we have found the contribution to the $\bar{z}...\bar{z}$ component multiplied with lowest derivative on $\Lambda^{(s)}$ due to definition of trace for generators $V_n^s$ \cite{Gaberdiel:2011wb}
\small
\begin{equation}
tr\left( V_m^s V_n^t \right)=N_s\frac{(-1)^{s-m-1}}{(2s-2)!}\Gamma(s+m)\Gamma(s-m)\delta^{st}\delta_{m,-n}. \label{trace}
\end{equation}
\normalsize
for 
\begin{equation}
N_s\equiv \frac{3\cdot 4^{s-3}\sqrt{\pi}q^{2s-4}\Gamma(s)}{(\lambda^2-1)\Gamma(s+\frac{1}{2})}(1-\lambda)_{s-1}(1+\lambda)_{s-1}
\end{equation}
and $(a)_n=\frac{\Gamma(a+n)}{\Gamma(a)}$ ascending Pochhammer symbol. The overall constant is set to \begin{equation}tr(V_1^2V_{-1}^2)=-1.
\end{equation}

Let us go back to $\phi_{\bar{z}....\bar{z}}$ component. The star product $e_{\bar{z}}\star...\star e_{\bar{z}}$ will contribute with $\frac{1}{2^{s-1}}e^{(s-1)\rho}V^{s}_{-(s-1)}$ if we consider as explained above the lowest derivative on $\Lambda^{(s)}$. We can denote this as 
\begin{align}
e_{\bar{z}}\star....\star e_{\bar{z}}(V_{-1}^2\star ....\star V_{-1}^2)=\frac{1}{2^{s-1}}e^{(s-1)\rho}V^{s}_{-(s-1)}.\end{align} The $\tilde{E}_{\bar{z}_s}=\tilde{A}_{\bar{z}_s}-\tilde{\bar{A}}_{\bar{z}_s}$ needs to be able to satisfy the conditions of the trace (\ref{trace}) in star multiplication with $e_{\bar{z}}\star...\star e_{\bar{z}}$, the only HS generator that  contributes is $V^s_{s-1}$ generator. When we gauge the field $A_{\bar{\mu}_s}$, $d\bar{z}$ component appears in $d\Lambda$ while $A_{AdS}$ and $[A_{AdS},\Lambda]_{\star}$ do not have $d\bar{z}$ component. The $\tilde{\bar{A}}_{\bar{z}_s}$ has $d\bar{z}$ component that comes from $\bar{A}_{AdS}$ part and it is $e^{\rho}V_{-1}^2d\bar{z}$. This however will not appear with the right number of derivatives on $\Lambda$. Since we have chosen $\Lambda$ to be chiral and $\bar{\Lambda}=0$, that was the only contribution from $\tilde{\bar{A}}_{\bar{z}}$. 

Altogether, we can write $\phi_{\bar{z}...\bar{z}}$ component for the $\bar{\partial}\Lambda^{(s)}$ derivative as
\small
\begin{align}
\phi_{\bar{z}....\bar{z}}|_{\bar{\partial}\Lambda^{(s)}}&=tr\left[\frac{1}{2^{s-1}}e^{(s-1)}V^s_{-(s-1)}\star V^s_{s-1}e^{(s-1)\rho}\bar{\partial}\Lambda^{(s)}(z,\bar{z})\right]\\
&=\frac{1}{2^{s-1}}e^{2(s-1)\rho}\bar{\partial}\Lambda^{(s)} N_s.
\end{align}
\normalsize
Inserting the normalisation $N_s$ we obtain
\begin{align}\nonumber
\phi_{\bar{z}...\bar{z}}|_{\bar{\partial}\Lambda^{(s)}}&=\frac{1}{2^{s-1}}e^{2(s-1)\rho}\bar{\partial}\Lambda^{(s)}\\&\times
\frac{3\cdot 4 \sqrt{\pi}4^{4-2s}\Gamma(s)\Gamma(s+\lambda)\Gamma(s-\lambda)}{(\lambda^2-1)\Gamma(s+\frac{1}{2})\Gamma(1-\lambda)\Gamma(1+\lambda)}.\label{phibarz}
\end{align}

The expression $\phi_{\bar{z}...\bar{z}}$ we want to compare with expression (\ref{boxeom}) for highest derivative on $C_0^1$ and $\bar{\partial}\Lambda^{(s)}$. 
In the computation of the vertex this would be a term 
\begin{equation}
\phi^{z...z}\mathcal{\phi}\nabla_z...\nabla_{z}\mathcal{\phi}
\end{equation}
for $\phi^{z...z}$ higher spin field with $s$ indices and $\mathcal{\phi}$ scalar field. Raising indices 
 contributes with a factor $2^se^{-2s\rho}$, so that the field $\phi^{z...z}$ becomes
\begin{align}
\phi^{z...z}&=\frac{1}{2}e^{-2\rho}\bar{\partial}{\Lambda}^{(s)}3\cdot4^{4-2s}\nonumber \\ &\times
\frac{\Gamma(s)\Gamma(s+\lambda)\Gamma(s-\lambda)}{(\lambda^2-1)\Gamma(s+\frac{1}{2})\Gamma(1-\lambda)\Gamma(1+\lambda)}.\label{phiz}
\end{align}
When  we take the ratio with $\Box_{KG}|_{\text{highest number of derivatives}(\delta C_0^1)|_{\bar{\partial}\Lambda}}=(-1)^s4e^{-s\rho}\bar{\partial}\Lambda^{(s)}\partial^sC_0^1$ we get (schematically written)

\begin{align}
&\frac{\phi^{z...z}|_{\bar{\partial}\Lambda^{(s)}}}{\Box_{KG}|_{\text{highest number of derivatives}(\delta C)_0^1|_{\bar{\partial}\Lambda^{(s)}}}}\nonumber\\&=(-1)^s\frac{1}{2}3\sqrt{\pi}\frac{4^{4-2s}\Gamma(s)\Gamma(s+\lambda)\Gamma(s-\lambda)}{(\lambda^2-1)\Gamma(s+\frac{1}{2})\Gamma(1-\lambda)\Gamma(1+\lambda)}.\label{phizbox}
\end{align}
which taking into account the normalisation gives the coupling for the 00s three point function.

\section{Conclusion and Outline}

We have considered the three-point coupling using metric-like formation to express the higher spin field and using the linearised Vasiliev's equations of motion.  The obtained result can also be verified using the alternative methods, for example following the procedure by \cite{Gaberdiel:2010pz}. 
The generalisation of the result to higher point functions would be non-trivial since in order to compute higher order vertices, one would have to consider perturbations around the background AdS field with higher spin fields up to that required higher order. 

\section{Acknowledgements} 
I would like to thank Stefan Fredenhagen for guidance and discussions, Arkady Tseytlin, and Jan Rosseel for discussions. The work was made possible by the hospitality of University of Vienna, and Imperial College London, and supported by the project No. J 4129-N27 of the Austrian Science Fund (FWF) in the framework of Erwin-Schr\"odinger Program, and by the grant ST/P000762/1 of Science and Technology Facilities Council (STFC).

\bibliographystyle{apsrev}
\bibliography{bibliot}

\end{document}